# Semantic contrastive learning for orthogonal X-ray computed tomography reconstruction

Jiashu Dong[1,2], Jiabing Xiang[1,2], Lisheng Geng[2], Suqing Tian[3], and Wei Zhao[1,2,4,*]

[1]Hangzhou International Innovation Institute, Beihang University, Hangzhou, China
[2]School of Physics, Beihang University, Beijing, China
[3]Department of Radiation Oncology, Peking University Third Hospital, Beijing
[4]Tianmushan Laboratory, Hangzhou, China

**Abstract** X-ray computed tomography (CT) is widely used in medical imaging, with sparse-view reconstruction offering an effective way to reduce radiation dose. However, ill-posed conditions often result in severe streak artifacts. Recent advances in deep learning-based methods have improved reconstruction quality, but challenges still remain. To address these challenges, we propose a novel semantic feature contrastive learning loss function that evaluates semantic similarity in high-level latent spaces and anatomical similarity in shallow latent spaces. Our approach utilizes a three-stage U-Net-based architecture: one for coarse reconstruction, one for detail refinement, and one for semantic similarity measurement. Tests on a chest dataset with orthogonal projections demonstrate that our method achieves superior reconstruction quality and faster processing compared to other algorithms. The results show significant improvements in image quality while maintaining low computational complexity, making it a practical solution for orthogonal CT reconstruction.

## 1 Introduction

Computed tomography (CT) is one of the most widely used medical imaging modalities, providing detailed internal structural information. Traditional reconstruction algorithms, such as filtered back-projection (FBP) and iterative reconstruction, require the sampling of X-ray projections to meet the Shannon-Nyquist theorem, imposing a practically achievable limit in imaging time. Additionally, acquiring a large number of projections increases the radiation dose, raising safety concerns for patients. Reducing the number of projections has become a common strategy to address these issues, yet it introduces new challenges for reconstruction algorithms, especially in ultra-sparse orthogonal projection scenarios where the problem is even more ill-posed.

In recent years, data-driven deep learning methods have demonstrated remarkable success in tackling ill-posed problems. Generative adversarial networks (GANs) have been applied to orthogonal projection reconstruction[1,2], leveraging their ability to generate projections from alternative viewpoints or directly reconstruct 3D images. Furthermore, the adoption of transformers as backbone networks in GAN architectures has improved global feature extraction[3]. However, GANs face inherent limitations, including training instability, mode collapse, and the generation of hallucinated features. Diffusion models, the latest advancements in generative modeling, produce high-quality, semantically accurate reconstructions through iterative denoising[4,5]. Nevertheless, these models suffer from slow training convergence and inference processes, resulting in significant computational overhead.

To address these limitations in orthogonal CT reconstruction, we propose a novel feature-based semantic contrastive learning loss function[6,7]. This loss function measures semantic similarity in high-dimensional latent spaces and anatomical feature similarity in low-dimensional spaces, ensuring robust and accurate reconstruction. By integrating semantic feature regularization into the GAN framework, we enhance training stability, mitigate mode collapse, and reduce hallucinated features. Our streamlined network architecture consists of three U-Nets: one for coarse reconstruction, one for detail refinement, and one for semantic similarity assessment. During inference, only two U-Nets are required, significantly reducing computational complexity and accelerating processing. This efficient design offers a practical solution for orthogonal CT reconstruction, providing high-quality results with reduced hardware and computational demands.

## 2 Materials and Methods

### 2.1 Network Overview

Figure 1(a) illustrates our workflow, which follows a coarse-to-fine training framework for orthogonal CT reconstruction. In the first stage, orthogonal projections are back-projected using geometric relationships to obtain an initial 3D reconstructed image. This image is fed into a 3D U-Net, yielding a coarse 3D image with basic anatomical structures. The mean squared error (MSE) between the output and the original CT image is used to optimize the network. At this stage, the reconstructed image has smooth edges, low contrast, and limited detail.

To enhance details and improve contrast, we introduce an image refinement network. Each slice of the 3D image is processed through a 2D U-Net for detail enhancement. A hybrid loss function is employed, combining the L1 loss $\mathcal{L}1$, VGG perceptual loss $\mathcal{L}_{vgg}$, adversarial loss $\mathcal{L}_{disc}$, and semantic contrastive learning loss $\mathcal{L}_{contrast}$. The perceptual loss captures structural and textural accuracy, the adversarial loss enhances fine details, and the semantic contrastive loss prevents hallucinatory artifacts by constraining detail generation.

Our architecture employs two five-level U-Net models for the coarse and refinement stages, each using an encoder-decoder structure. Encoders consist of two convolutional blocks followed by 2×2×2 max pooling for downsampling. Each block includes a 3×3×3 convolutional layer, instance normalization, and a Leaky ReLU activation function. Feature dimensions progressively increase at each level: 64, 128, 256, 512, and 1024. The decoder mirrors this structure, with 2× upsampling via transposed convolutions and similar convolutional blocks. This symmetric design facilitates effective feature extraction and reconstruction, enabling the model to recover both coarse structures and fine image details.

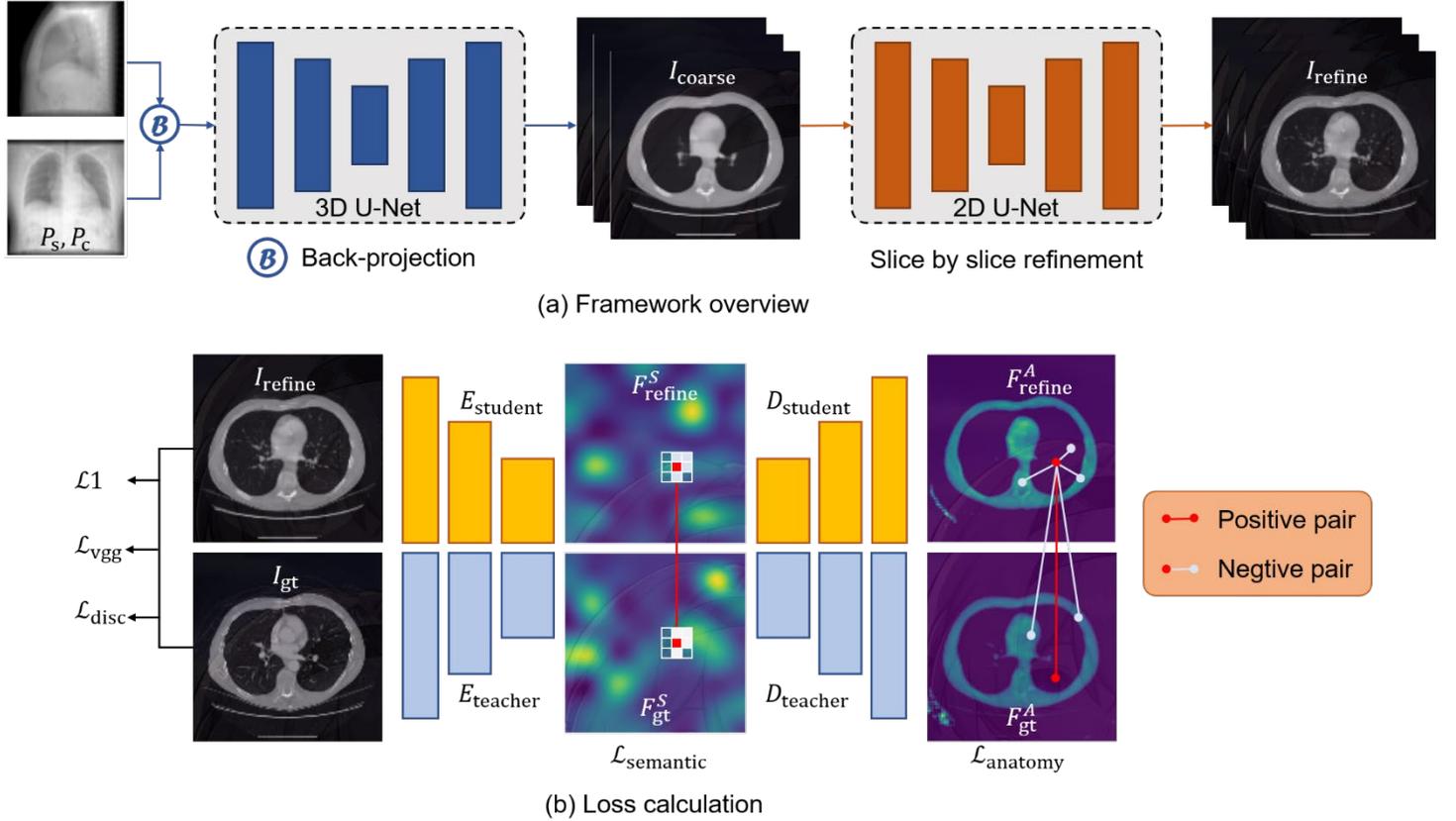

**Figure 1:** Diagrams of our framework and loss calculation.

## 2.2 Semantic Contrasive Learning

Figure 1(b) illustrates the calculation process for the semantic contrastive learning loss function during the image refinement stage. The contrastive learning network is based on a 2D U-Net but modified for this task: the upsampling operation is replaced with linear interpolation, and the final convolutional layer for output size adaptation is removed to better preserve features in the latent space. The network employs a self-distillation approach for parameter updates. The student model, with learnable parameters, generates semantic and anatomical features for the refined image. The teacher model, whose parameters is derived as an exponential moving average of the student model's parameters, generates semantic and anatomical features for the original image. The goal is to ensure that the representations of the refined and original images at corresponding positions are identical.

To achieve consistency for similar structures and differentiation for distinct ones, we employ specific sampling strategies for positive and negative samples based on the two feature types. For high-level semantic features, the pixels at the same position serve as a positive sample. Additionally, cosine similarity is calculated between the center pixel and its neighboring pixels, and the top $N_{pos}^S$ most similar pixels are selected to compute MSE as the semantic loss.

$$\mathcal{L}_{semantic} = \sum_{i \in N_{pos}^S} \| f_{refine,i}^S - f_{gt,i}^S \|_2$$

Where $f_{refine}^S$ and $f_{gt}^S$ represent the feature vector projections output by the student network and teacher network encoders, respectively. This ensures that anatomically similar structures, which are typically located nearby, are appropriately aligned. For low-level anatomical features, the network distinguishes subtle differences in structure by treating the most similar neighboring pixels as negative samples. Cosine similarity is again computed between the center pixel and its neighbors, and the top $N_{neg}^A$ most similar pixels are selected to calculate the InfoNCE loss.

$$\mathcal{L}_{anatomy} = -\sum_{i \in N_{pos}^A} \log \frac{\exp(f_{refine,i}^A \cdot f_{gt,i}^A / \tau)}{\exp(f_{refine,i}^A \cdot f_{gt,i}^A / \tau) + \sum_{j \in N_{neg}^A} \exp(f_{refine,i}^A \cdot f_{gt,j}^A / \tau)}$$

Where $f_{refine}^A$ and $f_{gt}^A$ represent the feature vector projections output by the student network and teacher network decoders, respectively. And τ is the temperature coefficient. This approach encourages the network to learn fine-grained distinctions in anatomical structures, enhancing the quality of the reconstruction.

### 2.3 Dataset and Experiment

We evaluated our model using the LIDC-IDRI dataset, which comprises 1,018 thoracic CT scans intended for diagnostic and lung cancer screening, sponsored by the National Cancer Institute (NCI) of the United States. After downloading the dataset, 1,009 CT scans were confirmed to be usable, as 9 scans were either corrupted or failed to download. The available images were randomly divided into a training set with 808 samples and a test set with 201 samples. Before training, the 3D volumetric images of the CT dataset are resampled to dimensions of 128×128×128 with a voxel spacing of 2.8mm × 2.8mm × 2.8mm.

Our model was implemented using PyTorch 2.1.2 and trained on an NVIDIA RTX 3090 GPU with 24 GB of memory. The training process followed a two-stage framework. In the first stage, the model was trained for 220 epochs using the AdamW optimizer with an initial learning rate of 0.0002, which decayed to 1e-6 following a cosine schedule. In the second stage, the model was trained for 100 epochs using the AdamW optimizer with an initial learning rate of 0.0001, also decayed to 1e-6 with a cosine schedule. To validate the effectiveness of our method, we reproduced and compared it against three deep learning-based approaches: 2D CNN, 3D U-Net, and X2CT-GAN.

### 2.4 Evaluation Metrics

To quantitatively assess the quality of the reconstructed images, we use several evaluation metrics, including mean absolute error (MAE), peak signal-to-noise ratio (PSNR), structural similarity index (SSIM), visual information fidelity (VIF), perceptual similarity (LPIPS) and Dice coefficient (DICE). The Dice coefficient is commonly used in image segmentation tasks to evaluate the overlap between predicted and ground truth contour structures. To assess the practicality of the reconstructed images, we segment the lung region using thresholding techniques and then compute the Dice coefficient against the ground truth. This ensures that our method is evaluated not only for reconstruction quality but also for its relevance to downstream tasks like lung region segmentation.

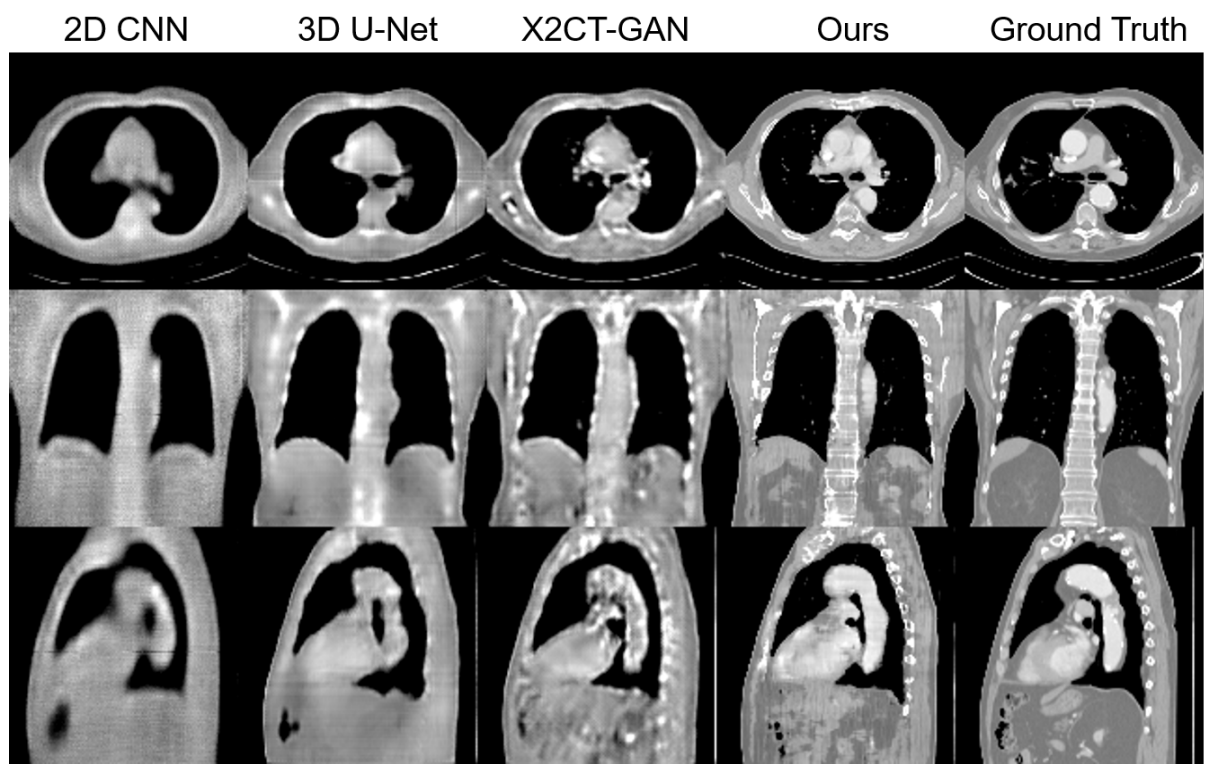

**Figure 2:** Comparison of the quality of reconstruction images. Displayed window is [-500, 800] HU.

## 3 Results

Figure 2 illustrates the reconstruction results of different networks, with the displayed window set to [-500, 800] HU. The 2D CNN interprets the channel dimension of extracted features as the Z-axis of the reconstructed image, using the same convolution kernel across different positions. This method fails to accurately capture body margins. The 3D U-Net, which employs 3D convolution layers, partially mitigates the limitations of 2D convolution networks by processing features across different regions. However, it still struggles with misaligned contours of critical internal structures and produces blurry interiors. Reconstruction results using X2CT-GAN can roughly identify the positions of the lungs and spine. However, it also introduces significant hallucinatory artifacts which are often observed in GAN-based methods, particularly in soft tissue areas, and yield unrealistic HU values near air pockets around bones. In contrast, our proposed semantic contrastive learning loss function effectively addresses these shortcomings. The reconstructed spine appears significantly clearer, with even substructures being distinguishable. Additionally, the shapes and textures of the lungs, heart, and other soft tissues are well-preserved, resulting in superior overall reconstruction quality.

Table 1: Quantitative analysis between different methods

| Metric | 2D CNN | 3D U-Net | X2CT-GAN | Ours |
|---|---|---|---|---|
| MAE↓ | 102.25 | 90.90 | 85.38 | **75.30** |
| PSNR↑ | 23.77 | 24.48 | 25.29 | **25.53** |
| SSIM↑ | 0.606 | 0.663 | 0.712 | **0.734** |
| LPIPS↓ | 1.092 | 0.887 | 0.813 | **0.595** |
| VIF↑ | 0.445 | 0.572 | 0.639 | **0.696** |
| DICE↑ | 0.749 | 0.646 | 0.788 | **0.875** |

Quantitative evaluation in Table 1 shows that our proposed model outperforms all other methods across all metrics, with substantial improvements in LPIPS, VIF, and DICE scores. These results demonstrate that our model better preserves texture and organ shapes while retaining more structural and semantic information compared to the original images.

## 4 Discussion

Orthogonal projection reconstruction of CT images is an extremely ill-posed inverse problem, offering significant advantages in rapid imaging and radiation dose reduction. However, designing effective reconstruction algorithms for this task has long been a challenge. Current GAN-based methods are prone to hallucinated artifacts, compromising their reliability, while diffusion model-based methods, despite their high-quality outputs, are hindered by slow reconstruction speeds, making them unsuitable for real-time applications.

Our proposed method addresses these limitations by employing a streamlined architecture that uses three U-Nets during training and only two during inference. The inference time is below 0.4s that ensures faster processing and reduces computational complexity. A key innovation is the integration of a semantic contrastive learning loss function during training, which guides the network to generate anatomical structures consistent with the original images. This semantic regularization effectively minimizes hallucinated artifacts within the GAN framework. The reconstructed images exhibit clear and accurate contours, preserving critical structures and textures. Both visual and quantitative evaluations demonstrate that our method achieves optimal reconstruction results.

Beyond standard imaging tasks, the potential applications of this method are vast. The ability to provide high-quality reconstructions with clear contours in critical regions can significantly reduce positioning time before radiotherapy, improving treatment precision and patient throughput. Furthermore, this approach is highly suited for real-time volumetric imaging during CyberKnife beam delivery, where rapid, accurate imaging is essential for real-time tumor tracking and adaptive radiotherapy. Additionally, the speed and quality of our reconstruction framework make it a promising tool in other time-sensitive medical applications, such as emergency diagnostics, intraoperative imaging, and portable CT scanners for remote or resource-limited settings.

## 5 Conclusion

In this study, we propose a coarse-to-fine orthogonal projection reconstruction method, which incorporates semantic information consistency regularization within the GAN framework, significantly addressing the issue of hallucinated artifacts. This method is fast in reconstruction, generate superior images, and shows promising potential for practical applications.

## Acknowledgments

This work was supported in part by the Natural Science Foundation of Zhejiang Province, [Grant/Award Number: LZ23A050002]; National Natural Science Foundation of China, [Grant/Award Number: 12175012]; the '111' center (B20065) and the Fundamental Research Funds for the Central Universities, and a grant from Xiaomi Scholarship.

## References

[1] Ying X, Guo H, Ma K, et al. X2CT-GAN: reconstructing CT from biplanar X-rays with generative adversarial networks[C]//Proceedings of the IEEE/CVF conference on computer vision and pattern recognition. 2019: 10619-10628.

[2] Gao Y, Tang H, Ge R, et al. 3DSRNet: 3D Spine Reconstruction Network Using 2D Orthogonal X-ray Images Based on Deep Learning[J]. IEEE Transactions on Instrumentation and Measurement, 2023.

[3] Zhang C, Liu L, Dai J, et al. XTransCT: ultra-fast volumetric CT reconstruction using two orthogonal x-ray projections for image-guided radiation therapy via a transformer network[J]. Physics in Medicine & Biology, 2024, 69(8): 085010.

[4] Zhang J, Mao H, Wang X, et al. Wavelet-inspired multi-channel score-based model for limited-angle CT reconstruction[J]. IEEE Transactions on Medical Imaging, 2024.

[5] Huang B, Lu S, Zhang L, et al. One-Sample Diffusion Modeling in Projection Domain for Low-Dose CT Imaging[J]. IEEE Transactions on Radiation and Plasma Medical Sciences, 2024.

[6] Chen Z, Gao Q, Zhang Y, et al. Ascon: Anatomy-aware supervised contrastive learning framework for low-dose ct denoising[C]//International Conference on Medical Image Computing and Computer-Assisted Intervention. Cham: Springer Nature Switzerland, 2023: 355-365.

[7] Yan K, Cai J, Jin D, et al. SAM: Self-supervised learning of pixel-wise anatomical embeddings in radiological images[J]. IEEE Transactions on Medical Imaging, 2022, 41(10): 2658-2669.